\documentclass[twoside,twocolumn,english,superscriptaddress,nofootinbib,preprintnumbers, aps]{revtex4}
\usepackage{color}
\usepackage{amsmath}
\usepackage{graphicx}
\usepackage{amssymb}
\usepackage{esint}
\usepackage[hyperfootnotes=true,unicode=true, 
 bookmarks=true,bookmarksnumbered=false,bookmarksopen=false,
 breaklinks=false,pdfborder={0 0 1},backref=false,colorlinks=true]
 {hyperref}

\makeatletter
\@ifundefined{textcolor}{}
{%
 \definecolor{BLACK}{gray}{0}
 \definecolor{WHITE}{gray}{1}
 \definecolor{RED}{rgb}{1,0,0}
 \definecolor{GREEN}{rgb}{0,1,0}
 \definecolor{BLUE}{rgb}{0,0,1}
 \definecolor{CYAN}{cmyk}{1,0,0,0}
 \definecolor{MAGENTA}{cmyk}{0,1,0,0}
 \definecolor{YELLOW}{cmyk}{0,0,1,0}
 }

\usepackage{hyperref}
\hypersetup{
    colorlinks,%
    citecolor=blue,%
    filecolor=blue,%
    linkcolor=blue,%
    urlcolor=blue
}

\makeatother

\begin{document}

\preprint{CERN-PH-TH/2011-244}

\title{Price for Environmental Neutrino-Superluminality}

\author{Gia Dvali}
\email{georgi.dvali@cern.ch}
\affiliation{CERN, Theory Division, 1211 Geneva 23, Switzerland} 
\affiliation{ASC, Department f\"ur Physik, LMU M\"unchen, 
Theresienstr. 37, 80333 M\"unchen, Germany}  
\affiliation{MPI f\"ur Physik, F\"ohringer Ring 6, 80805 M\"unchen, Germany} 
\affiliation{CCPP, Department of Physics, New York University, 4 Washington Place, New York,
NY 10003, USA}

\author{Alexander Vikman}
\email{alexander.vikman@cern.ch}
\affiliation{CERN, Theory Division, 1211 Geneva 23, Switzerland}

\date{\today{}}

\begin{abstract}
We ask whether the recent  OPERA results on neutrino superluminality 
could be an environmental effect characteristic of  the local neighborhood of our planet, 
without the need of violation of  the Poincar\'e-invariance at a fundamental level.  We show, that model-indepenently, such a possibility implies the existence of new gravitational degrees of freedom. 
Namely, this explanation requires the existence of a new spin-2 field of a planetary  Compton 
wave-length that is  coupled to neutrinos and the rest of the matter asymmetrically, both in the magnitude and in the sign.   Sourced by the earth this field creates an effective metric  on which  neutrinos propagate superluminally, whereas other species are much less sensitive to the background.   
Such a setup, at an effective field theory level, passes all immediate phenomenological tests, but at the expense of sacrificing calculability for some of the phenomena that are under perturbative control in ordinary gravity. The natural prediction is an inevitable appearance of a testable long-range gravity-type fifth force. Despite phenomenological viability, the sign asymmetry of the coupling we identify as the main potential obstacle for a consistent  UV-completion. We also discuss the possible identification of this field with a Kaluza-Klein state of an extra dimension in which neutrino can propagate.
\end{abstract}

\maketitle

This note is inspired by recent  results by  OPERA \cite{OPERA:2011zb}  about possible evidence for superluminal propagation of neutrinos.   Needless to say, discovery of superluminality would require major rethinking of our understanding of principles of relativity. In this note we shall assume that OPERA results indeed point to superluminality  
of neutrinos and ask what minimal changes in the Standard Model physics could accommodate   
such a phenomenon.    
   
    An immediate challenge is to reconcile OPERA results with  the absence of analogous observations for superluminal propagation for supernova neutrinos. 
    
 One possible approach would be to suggest  violation of the Poincar\'e invariance at the fundamental level through some energy-dependent operators that would result in  superluminal propagation in an energy-dependent way, see e.g. \cite{Cacciapaglia:2011ax,OperaOperator}.  This is a logical possibility, but we shall take a different root. 
 
  We shall not postulate any violation of the Poincar\'e symmetry at the fundamental level. Instead, we shall 
ask whether the effect is environmental and takes place in the local neighborhood of the earth.
We shall show that such an explanation, under the assumption of a calculable weakly-coupled physics,  leads us to inevitability of  existence  of a new gravity-type force, asymmetrically-coupled to neutrinos and to the heavy  matter, such as nucleons.     
 
  We shall structure our discussion in the following way.   We shall first show that the 
  environmental  superluminality can be explained by introduction of the above-mentioned 
 new  gravity-type force. Then we will prove that this is the only possible environmental explanation under the 
 assumption of the weak coupling and calculability of the neutrino propagation. This new force in its calculability range is compatible with all the immediate phenomenological constraints. But, the phenomenological price to pay is a very low strong coupling scale, which sacrifices
calculability for some systems for which the ordinary gravity would be weakly coupled and is under control.
 It is extremely important that appearance of such a low cutoff is an inevitable consequence of any
environmental explanation of OPERA results.
 
   Thus, let us accomplish the first goal  by postulating existence of a new light bosonic degree of freedom  with the Compton wavelength of the order of planetary distances. 
 The role  of this degree of freedom is simple.  We assume that this field is sourced by earth and creates a classical background to which neutrino is coupled. 
 Neutrino then propagates through an effective metric  that speeds  it up. 
 
  For concreteness, we shall illustrate this idea on an example of a new massive spin-2 degree of freedom, $h_{\mu\nu}$,   coupled to neutrino in the following way,  
 \begin{equation}
    (\eta_{\mu\nu} \, + \, {h_{\mu\nu} \over M_*} ) \, \bar{\nu} \gamma^{\mu}\partial^{\nu} \nu \, . 
    \label{coupling} 
    \end{equation}
    For simplicity of presentation,  we shall treat neutrinos as massless. We view the above expression as an effective low energy coupling in which all the heavy weak-scale physics has been integrated out. As a result,  neutrino sees the following effective metric \footnote{For neutrino this metric plays a role of the \emph{contravariant} metric in standard GR, while signal propagates in the \emph{covariant} metric which is inverse to the one in \eqref{metric}. Throughout the paper we raise and lower indices with $\eta^{\mu\nu}$ and $\eta_{\mu\nu}$ respectively.}
  \begin{equation}
    g_{\mu\nu}^{(\nu)} \, = \, \eta_{\mu\nu} \, + \, {h_{\mu\nu} \over M_*}  \, . 
    \label{metric} 
    \end{equation}
  The scale $M_*$ sets the strength of the coupling.  Thus, the interaction of $h_{\mu\nu}$ with 
  neutrino is similar to linear gravity. However,  for us $h_{\mu\nu}$ is just another massive spin-2 
  field, not necessarily of any geometric origin.     
   Correspondingly,  the scale $M_*$ is not the Planck mass, and its value will be constrained below from the OPERA data. 
    On a non-trivial background with $h_{\mu\nu}$ field the above coupling effectively amounts to 
   changing the anti-commutation relation of effective gamma matrices to a new metric.  
   
      We now have to specify the coupling of $h_{\mu\nu}$ to other Standard Model particles. 
 Of main importance is the  coupling with the species that  give dominant contribution into the 
 earth's mass, such as, nucleons. Since we are not making any assumptions about the gravitational origin of  $h_{\mu\nu}$,  its couplings to the latter states do not have to obey the equivalence principle.   
 The simplest possibility, however, is when the coupling to the rest of the species is universal and is through energy-momentum tensors, 
 \begin{equation}
     {h_{\mu\nu} \over M} \, T^{\mu\nu} \, ,
    \label{coupling1} 
    \end{equation}
  where $T_{\mu\nu}$ should not include neutrino.  Since in our analysis we shall work at 
  the level of very low energy effective theory,  $T_{\mu\nu}$ can be directly taken to be an effective 
  energy momentum tensor of the earth. 
  
   In order to complete our analysis, we will need a Lagrangian for $h_{\mu\nu}$.   For the range 
   of energies and distances of our interests the linearized analysis will be fully sufficient and reliable.  Therefore, we shall restrict ourselves with the linear action, which is uniquely fixed to be 
  of the Pauli-Fierz form, 
  \begin{equation}
    h^{\mu\nu}{\mathcal E}  h_{\mu\nu} \, + \, m^2  (h_{\mu\nu} h^{\mu\nu} \, - h_{\mu}^{\mu} h_{\nu}^{\nu} ) \, ,  
    \label{PF} 
    \end{equation}
 where  ${\mathcal E}h_{\mu\nu}$ is the linearized Einstein's tensor. 
 
   Non-linear interactions shall play no role in our analysis. We are fully aware of subtleties 
   of non-linearities,   since they usually result into low cutoffs. 
  It is not our goal to extend the theory beyond these cutoffs, and we shall safely stay below it. Even assuming a most conservative 
   case, the scale of non-linearities (a so-called Vainshtein scale \cite{Vainshtein:1972sx}) for our choice of parameters dictated by OPERA, appears way beyond the range of our interest. 
  As a result, we can perform a fully reliable computation in a linear regime.   
    
  Thus, the effective  Lagrangian we work with represents the sum of the three terms 
 given in equations (\ref{coupling}), (\ref{coupling1}) and (\ref{PF}). 
    As we shall see, in order to explain OPERA results, the scales $M$ and $M_*$ must be above and below the Planck mass respectively.  This choice results in the following situation. 
  Mass of the earth sources $h_{\mu\nu}$ and creates a local classical field. 
  This field will have a negligible effect on a local gravitational background seen by all the particles except neutrino. The latter shall feel the $h_{\mu\nu}$ background  much stronger and as a result
 become slightly superluminal.  
     
     In order to see this, let us find a static background of $h_{\mu\nu}$ created by the earth.
   This is provided by the solution of the linearized equation 
    \begin{equation}
   (- \Delta \, + \, m^2)  h_{\mu\nu} \, = \, {1 \over M} (T_{\mu\nu} \, - {1\over 3} (\eta_{\mu\nu} \, + \, {\partial_{\mu}\partial_{\nu} \over m^2})\,  T)  \, ,  
    \label{line} 
    \end{equation}
in which $T_{\mu\nu}$ is taken as a non-relativistic spherical source of the earth's mass,  $M_E$. 
 The result for time and space components  at distance $r \, \ll \, m^{-1}$ is,  
  \begin{equation}
  h_{00} \, = \, {2\over 12\pi} \eta_{00}  {M_E \over M r} \, , ~~~ \,      
   h_{ij} \, = \, - \, {1\over 12\pi}  \eta_{ij}  {M_E \over M r} \,  ,      
\label{solution}
\end{equation}
where the contribution proportional to total derivatives has been neglected due to conservation 
of the probe neutrino source. 
Correspondingly the effective metric in which neutrino propagates is 
  \begin{equation}
  g_{00}^{(\nu)} \, = \, ((1 - {1\over 3}\epsilon) \,  + \,  \epsilon)  \eta_{00}  \, , ~~~ \,      
   g_{ij}^{(\nu)} \, = \,  (1 - {1\over 3}\epsilon)  \eta_{ij}  \,  ,      
\label{metricfinal}
\end{equation}
where we have introduced a notation\footnote{Unless the units explicitly appear, we work in the reduced Planck units where 
$M_P \equiv \left(8\pi G\right)^{-1/2}=1$.}
\begin{equation}
\epsilon \equiv  \frac{M_E}{4 \pi M_*M r} \, .
\label{epsilon}
\end{equation}
Clearly the property of superluminality is determined by the sign of $\epsilon$, which depends 
on the relative sign of $M$ and $M_*$.  When the sign is negative, $\epsilon \, <\, 0$, the propagation is superluminal. 

The OPERA results correspond to $\epsilon \, \sim \, 10^{-5}$.  Then taking  the distance of the order 
of the earth-radius $r \, \sim \, 10^{8} \, \text{cm}$, we get that the OPERA observation can be reproduced by 
\begin{equation}
M_*M\, \sim \, 10^{-4} M_P^2 \, . 
\label{mainrealation}
\end{equation}
  An independent important constraint on the scale $M$ is coming from the absence of 
  any observable long-range fifth force of gravity-type.  Depending on the precise nature of couplings  this fact implies the constraint on $M$ in a wide window, $M^2/M_P^2 \, > \, 10^{4} \, - 10^{12}$. This bound follows from applying the experimental fifth force bounds \cite{Talmadge:1988qz} to the case of additional graviton(s) with Compton wave-length of the earth's radius derived in details in \cite {Dvali:2001gm}, where such gravitons  where motivated by the studies of earth-size extra dimension.  These results  can be directly applied to our case. The upper edge of the interval would take place in case of maximal violation of the equivalence principle. 
 
  Assuming universality, and combining the two bounds, we get, $M_* \sim 10^{-6} M_P$ and 
  $M\sim 10^2 M_P$.  
  
   Coming back to the consistency of our estimate, let us note that 
  for such values of the parameters, even in the worst possible scenario (in which no weakly-coupled completion exists before the strong coupling scale) the upper bound on the Vainshtein's  radius (distance at which non-linearities become important)  is at 
  $R_V \,  \sim \, ((M_E/M^2) m^{-4})^{1/5}$.   For  $M\sim 10^2 M_P$ and $m^{-1} \sim 10^8 \, \text{cm}$ 
   we get $R_V  \,  \sim  \,  10^{5} \, \text{cm}$,   which is way inside the earth's radius. 
 Thus  for our purposes, the  linear regime is a very good approximation and can be trusted.   
 
   We see that in order  to account for the OPERA result, the hierarchy of couplings can be  relatively mild.  Note,  that such a  hierarchy is radiatively stable, since $1/M_*$ coupling of  $h_{\mu\nu}$ to hadrons will be communicated only at the two-loop level, being suppressed by 
powers of the weak coupling constant.\footnote{Of course, for generic values of parameters  the radiatively-denegated equivalence-violating couplings have to be properly tuned.} Notice  that in this natural window, the fifth force is close to its experimental limits and is potentially testable. Thus, in a framework in which neutrino superluminality  is environmentally-induced, the 
 fifth force is a natural consequence of the scenario.

 The scale $M_*$ is further constrained by astrophysical and cosmological bounds coming from the  star-cooling, and the Big Bang Nucleosynthesis (BBN), see \cite{Dvali:2001gm}.   The requirement that production of $h_{\mu\nu}$ in the stars 
  gives a negligible correction to the cooling rate, implies an approximate  bound, $M_* \,  > \, 10^{7-8}\, \text{GeV}$. 
   The cosmological bound is derived by requiring that the production rate of $h_{\mu\nu}$ during
   BBN,  $\Gamma \, \sim \, T^3_N /M_*^2$,   is  subdominant to the expansion rate of the Universe, 
  $H \, \sim \ T^2_N /M_P$. This implies the bound, $M_* \, > \, \sqrt{M_PT_N}\, \sim \, 10^{7-8}\, \text{GeV}$, 
  where we have taken BBN temperature to be $T_N \sim 10 \, \text{MeV}$. 
 Interestingly, the two bounds are very close, which is similar to situation \cite{ArkaniHamed:1998nn} with analogous bounds on Kaluza-Klein graviton production in large extra dimensional scenario \cite{ArkaniHamed:1998rs}.   
Note that these bounds $M_* \, > \,  10^{8} \, \text{GeV}=10^{-10}M_P$ combined with \eqref{mainrealation} imply that $M\, < \, 10^{6} M_P$. Thus the window for $M_*$ is $10^{-10} M_P < M_* < 10^{-6}M_P$.

  Thus,  we have shown that additional asymmetric gravity-type force can provide environmental  
 explanation of neutrino superluminality. We wish now to make a stronger statement and show that under the assumption of weak-coupling such a force is the only possibility. The argument  goes in the following way. The environmental explanation by default  implies that 
 superluminality results from an effective background metric  $g_{\mu\nu}^{(\nu)}$ caused by 
the environment.  Since the effect is small, this effective metric can be represented as a small deviation 
from the flat metric 
 \begin{equation}
 g_{\mu\nu}^{(\nu)} \, = \, \eta_{\mu\nu} \, + \, \delta g_{\mu\nu}^{(\nu)} \, . 
 \label{soft}
 \end{equation}
  Since by assumption the theory is in a weak-coupling regime and perturbations are short-range and local,  the effective metric perturbation can be expanded in terms of canonically normalized {\it massive} degrees of freedom characterized by representations of the Poincar\'e group. 
  The most general, up to two-derivative, linear expansion has the following form \footnote{Similarly to \cite{Gauthier:2009wc} we could include nonlinear terms like $\partial_{\mu}\phi\partial_{\nu}\phi$. However, these terms would lead to a higher order interaction and correspondingly to a lower strong-coupling scale.}
  \begin{equation}
  \delta g_{\mu\nu}^{(\nu)} \,  = \,  {h_{\mu\nu}  \over M_*} \, +  \, \eta_{\mu\nu} {\phi \over M_0} \, + \, 
  {\partial_{\mu}\partial_{\nu} \phi  \over M_0^{'3}} \, + \, \frac{\partial_{\mu}A_{\nu}  + \partial_{\nu}A_{\mu}}{M_1^2} \ , 
 \label{expansion}
 \end{equation}
  where $h_{\mu\nu}$, $\phi$ and $A_{\mu}$ contain massive spin-2, spin-0 and spin-1 
  degrees of freedom respectively and $M_0, M_0', M_1$ are some mass scales. 
  The degrees of freedom that appear with derivatives do not contribute to couplings with the 
  conserved source at the linear level.  This leaves us with spin-2 and non-derivatively coupled spin-0 only. However, the coupling of spin-0 is  proportional to $\eta_{\mu\nu}$, and thus, at the linear level no superluminality can be induced by coupling to $\phi$. 
This leaves us with the above-discussed spin-2 option.   
 \\
 \\
   An important open question is of course existence of a sensible UV-completion for such class of theories, cf. \cite{Adams:2006sv}.   We have seen that  accommodation of neutrino superluminallity imposes a non-universal sign coupling  of a new spin-2 state.  It is unclear whether such non-universally-coupled spin-2 states can be embedded in a consistent microscopic theory. 
We have not addressed this issue in the present work, but we have to note, that should 
neutrinos be experimentally proven to be superluminal,   the analogous question  with be unavoidable  for any effective theory that addresses such superluminality.\footnote{One can turn the above argument around, and use the absence of UV-completion as the evidence against superluminal propagation.  But then there would be no reason to write this note to start with.}
Note that superluminality does not necessary lead to a breakdown of such a basic notion as causality, see e.g. \cite{Babichev:2007dw} cf. \cite{Adams:2006sv}
 
In connection with  UV-completion one can ask whether our $h_{\mu\nu}$ field can be identified 
 with a Kaluza-Klein excitation(s) of large extra dimension to which neutrino can propagate.\footnote{Apparent superluminality of neutrinos due to the propagation in an extra dimension was also considered in \cite{Pas:2005rb}. It can also happen that the speed of light in the bulk is larger than in our brane, see e.g. \cite{Kiritsis:1999tx}. Again, in 4d effective field theory language these options reduce to the scenario described above.} Such a setup was already suggested as a possible origin of small neutrino masses \cite{ArkaniHamed:1998vp}. This idea exploits  the  fact that the right handed neutrino is the only gauge-neutral particle and can naturally live in extra space, and thus share very weak-coupling properties with the graviton. 
In such a scenario neutrino naturally experiences different  couplings  with bulk gravity as compared to other standard model species.  It is tempting to identify our spin-2 field $h_{\mu\nu}$ with one of (or the entire tower) of massive bulk Kaluza-Klein states. Of course, correspondingly  the size of extra dimension has to be chosen to be comparable to  earth's radius,  perhaps along the lines of construction given in \cite{Dvali:2001gm}. 
The scale $M_*$ then must originate from the wave-function overlap integral between neutrino 
and Kaluza-Klein species.  Having this overlap the sign opposite to the ordinary graviton 
requires a very peculiar wave-function profiles, and currently we are not aware of any 
stable geometry of the extra dimensional space that could deliver it.  
This fact can be added  as one particular difficulty for UV-completion.   
So the question of such an embedding  will not  be answered in the present work. 
 \\
So far we have considered the coupling of the massive spin-2 field to the energy-momentum tensor of a free and masseless neutrino field. Due to the Standard Model interactions this incomplete energy-momentum tensor is not conserved. This non-conservation introduces derivative couplings of the longitudinal component $h^{\text{LL}}_{\mu\nu}$ of the massive spin-2 field:
\begin{equation}
h^{\text{LL}}_{\mu\nu}={\varphi} \, \eta_{\mu\nu} \, + \, {\partial_{\mu}\partial_{\nu} \varphi  \over m^{2}} \ ,
\end{equation}
to the $Z$ and $W^{\pm}$ bosons, neutrino $\nu$ and the corresponding lepton $\ell_{\nu}$, like e.g.:
\begin{equation}
\text{g} \, \frac{\partial_{\mu}\varphi  \,  \partial^{\mu}\bar{\nu} \, \gamma^{\alpha}\ell_{\nu} W_{\alpha}^{+}}{m^2 M_{*}} \ ,
\end{equation}
and
\begin{equation}
\frac{\text{g}}{\cos{\theta_{\text{w}}}} \, \frac{\partial_{\mu}\varphi  \, \partial^{\mu}\bar{\nu} \, \gamma^{\alpha}{\nu} Z_{\alpha}}{m^2 M_{*}} \ ,
\end{equation}
where g is the weak coupling constant and $\theta_{\text{w}}$ is the Weinberg angle. These derivative interactions do not change the neutrino front velocity because the gauge bosons have vanishing vacuum expectation value. However, these irrelevant operators introduce a strong-coupling scale 
\begin{equation}
\Lambda= \left(m^2 M_{*}\right)^{1/3}\ , 
\end{equation}
which happens to coincide with the scale of Dark Energy for our choice of parameters:
\begin{equation}\label{strong}
\Lambda \, = \, M_{P}\left( \frac {\ell_P}{r} \right)^{2/3} \left( \frac {M_{*}}{M_P} \right)^{1/3} \, \sim \,10^{-3} \, \text{eV}\ .
\end{equation}
After the first version of this note appeared on arXiv e-Print server the neutrino superluminality was, in particular, confronted with: i) too strong neutrino energy loss  due to bremsstrahlung of electron-positron pairs \cite{Cohen:2011hx}; and ii) the pion decay kinematics \cite{Cowsik:2011wv}. Unfortunately, in all these processes, at the interesting energy range, the scalar graviton $\varphi$  quanta  can not only be very efficiently emitted by neutrino but can also be absorbed from the condensate of $\varphi$ - background which is induced by the earth. The rather low strong-coupling scale \eqref{strong} for these decay and absorption  channels invalidates the perturbative calculations for the bremsstrahlung and requires a revision of the results on the pion decay kinematics. However, a detailed analysis of these effects goes beyond the purpose of this short note. Although our setup renders the effects of \cite{Cohen:2011hx} and \cite{Cowsik:2011wv}  un-calculable in weak-coupling, our analysis is complementary, since it opens up an universal gravitational side
of the problem.
\\
Here we would like to stress that, for any particle species, a non-fundamental modification of their front velocity (of the effective metric where they propagate) with respect to the speed of light (gravitational metric): a) can only occur in theories with irrelevant, nonrenormalizable operators and b) can only be caused by a spontaneous breaking of the Lorentz invariance. These irrelevant, nonrenormalizable operators introduce a strong-coupling scale. Thus any change of the front velocity is necessarily accompanied with a novel strong coupling scale which was not present in the standard model.
 \\
 \\
To conclude we have investigated an idea that superluminality of neutrinos can be a local environmental effect. We have shown that model-independently this would imply the existence of a new gravity-type field that is sourced by the earth and creates an effective superluminal metric for neutrinos. This follows from the uniqueness of the mode-expansion of effective metric perturbation on any asymptotically Poincar\'e-invariant background. We have seen that treated as an effective low energy field theory such a setup passes all the immediate tests and may avoid other, more involved, and more recent phenomenological constraints.  However, as we have demonstrated, the price to pay for the environmental neutrino superluminality is rather high. In particular, it necessarily includes: i) strongly-coupled physics on scales tremendously lower than a few TeV - the lowest cutoff scale for the currently known physics, ii) the sign-asymmetry in the coupling of massive graviton. We identify the latter fact as the main obstacle for a consistent UV-completion. While the too low strong-coupling scale results in the partial loss of calculability for a subsector of the standard model.
\\
\\
On phenomenological front this setup results into a natural prediction of a gravity-type force of an approximately planetary range. Without any proper adjustment, such a force is expected to be isotope-dependent, and thus, could be tested in precision gravitational experiments that look for equivalence-principle-violating forces.
\\
\acknowledgments{
We thank  Lasha Berezhiani, Fedor Bezrukov, Giga Chkareuli,  Cesar Gomez and Goran Senjanovic for discussions.  The work of G.D. was supported in part by Humboldt Foundation under Alexander von Humboldt Professorship,  by European Commission  under the ERC advanced grant 226371,  and  by the NSF grant PHY-0758032.}
 
\providecommand{\href}[2]{#2}\begingroup\raggedright\endgroup

\end{document}